# Improved Network Performance via Antagonism: From Synthetic Rescues to Multi-drug Combinations

Adilson E. Motter

Department of Physics and Astronomy and Northwestern Institute on Complex Systems, Northwestern University, Evanston, IL, USA

**Recent research shows that a faulty or sub-optimally operating metabolic network can often be rescued by the targeted *removal* of enzyme-coding genes—the exact opposite of what traditional gene therapy would suggest. Predictions go as far as to assert that certain gene knockouts can restore the growth of otherwise nonviable gene-deficient cells. Many questions follow from this discovery: What are the underlying mechanisms? How generalizable is this effect? What are the potential applications? Here, I will approach these questions from the perspective of compensatory perturbations on networks. Relations will be drawn between such *synthetic* rescues and naturally occurring cascades of reaction inactivation, as well as their analogues in physical and other biological networks. I will specially discuss how rescue interactions can lead to the rational design of antagonistic drug combinations that select against resistance and how they can illuminate medical research on cancer, antibiotics, and metabolic diseases.**

# Introduction

*"When the rats were deprived of a specific fat in their diet, their body cells compensated by overproducing it. (...) So, maybe, we could stop Lorenzo's body from producing saturated C24 and C26 by loading his diet with another kind of fat... you know, one that is less... is less harmful"*, said the character of Susan Sarandon in the true-story movie *Lorenzo's Oil*[1]. The movie portraits the discovery of a treatment for the inherited metabolic disorder adrenoleukodystrophy (ALD) by competitive inhibition of an enzyme that forms very long-chain fatty acids[2]. How many other diseases could be treated by "less harmful" (or not harmful at all) compensatory perturbations?

Potentially many more than previously anticipated. Contrary to what common wisdom may suggest, most cellular functions are carried out by the coordinated activity of multiple interacting elements, including genes, proteins and biochemical reactions[3,4]. A perturbation triggered by a genetic or epigenetic defect will often propagate through the cellular network causing the down-regulation of some reactions and up-regulation of others. But far from being a problem, the integrated nature of this system may hold the key for recovery. Indeed, while it remains generally unclear how global network properties are affected by local ones, recent progress has been made by inverting this perspective and seeking instead the conditions that we should impose on the local network structure and/or dynamics to generate a desired global collective behavior[5]. In the case of a defective cell, the desired behavior is the one that minimizes the impact of the defect. Such rescuing interventions can be argued to generally exist in systems that, like living cells, are governed by large networks that are *decentralized* and enjoy a certain level of *redundancy*.

This point is neatly illustrated by in silico studies of *Escherichia coli*'s metabolism. For *E. coli* fully adapted to arabinose minimal medium, for example, the knockout of gene *fbaA* is observed to be lethal, owing to the inability of the mutant strain to metabolize building blocks of the biomass, such as phenylalanine, tyrosine and L-lysine (see Fig. 1). But the shutdown of biomass production is accompanied by the activation of pathways, such as glyoxylate, that are predicted to be inactive for the mutant strain should it be able to operate in a state that maximizes biomass production. This observation suggests a possible recipe for the design of compensatory perturbations. Indeed, the inactivation of the glyoxylate pathway through the knockout of gene *aceA* is predicted to restore the organism's ability to produce biomass and hence grow (Fig. 1). This inactivation up-regulates the activity of reactions involved in the production of biomass precursors by globally rerouting fluxes to effectively bypass the defect caused by the knockout of gene *fbaA*. Because *fbaA*-deficient mutants are unable to reproduce—a prediction supported by experiments[6]—such flux changes would not occur spontaneously through adaptive evolution under the given arabionose minimal medium, making the proposed rescue all the more interesting. We refer to such knockout-based rescue interventions as *synthetic rescues*[5].

The fact that the metabolic network operates in a *decentralized* way (it does not have a central "controller") implies that it will generally respond to perturbations in a non-optimal manner with respect to any (natural or human-selected) objective function. Cells will, in particular, not grow as fast as they could under the constraints imposed by a perturbation even if the pre-perturbation state is optimal. The fact that the network is *redundant* (different metabolic states can often lead to comparable global objectives) implies that the activity of certain pathways, such as glyoxylate in the example above, can be constrained without compromising cellular function. More important, by doing so, fluxes can be forced to be routed through pathways that upon perturbation, such as the knockout of gene *fbaA*, are potentially more efficient for the cellular function of interest. Therefore, decentralization means that there is space for improvement, and redundancy that the response of the system can be in fact improved. These properties are actually common to many biological, physical and even social networks, rendering a significant level of generality to the concept underlying synthetic rescues. They have implications, for example, for the control of network-mediated failures ranging from blackouts and traffic jams to extinction cascades. In the specific context of cellular processes considered in this essay, they touch upon issues as diverse as epistasis[7], antagonistic drug interactions[8], gene dispensability[9], and the very notion of gene essentiality[10].

## Restraint-Based Control of Network Response

Synthetic rescues can be identified through a fairly general two-step procedure: first, identify the actual reaction fluxes and the reaction fluxes that correspond to the desired metabolic states (e.g., "optimal growth"); second, regard reactions with fluxes much larger than the desired ones as candidates for (total or partial) knockouts. The knockout of one or few such reactions, implemented through the knockout of the corresponding enzyme-coding genes, often brings the entire system closer to the desired state. This is precisely the criterion used to design the synthetic rescue of the *fbaA*-deficient mutant described above. Although this is admittedly a simplified description, it is not necessarily model-dependent as it allows for implementations based on experimental flux measurements using, for example, increasingly accurate $^{13}$C tracer techniques[11]. For instance, if the desired states are the ones that maximize the growth rate of a sub-optimally growing strain, candidate synthetic rescues can be identified by measuring the reaction fluxes of the strain and those of an adaptively evolved copy of the same strain. Perhaps the best way to appreciate this effect is, thus, by first examining what nature does to achieve similar goals.

Numerous metabolic reactions, such as those of the glyoxylate and Entner-Doudoroff pathways, are routinely observed to become dispensable or even incapable of carrying activity under steady genetic and environmental conditions. But recent experimental studies on *E. coli* have demonstrated that genetic[12] and environmental[13] perturbations are generally followed by the transient activation of a large number of latent pathways, i.e., pathways that, like glyoxylate in the example of Fig. 1, would not be recruited under unperturbed conditions. Growth, on the other hand, will tend to decrease in response to the same perturbation. The picture that emerges is thus the one in which a strain fully adapted to a certain environment will often experience a decrease in growth rate accompanied by a burst of reaction activity following a perturbation caused, say, by a gene knockout (see Fig. 2A, B). If the post-perturbation organisms (albeit less fit) remain able to grow, they may fully adapt to the new conditions after acquiring just a handful of (regulatory) mutations in the course of few hundreds or thousands of generations of adaptive evolution[14]. The final growth rate[15] and reaction activity[12] are often comparable to the original one, which suggests that the post-perturbation lulls are a consequence of suboptimal response as opposed to limitations inherent to the metabolic network.

This behavior has been further elucidated by the computational modeling of various single-cell organisms (see Box 1). It has been shown[16] that organisms evolved to maximize growth or one out of many other linear functions of metabolic fluxes, such as the sum of all fluxes, will necessarily activate a number of reactions that is just slightly larger than the number necessary for the organisms to grow at all[17,18]. It has also been shown[16] that typical suboptimal states necessarily recruit a much larger number of reactions. For *E. coli* fed glucose, for example, the number of active metabolic reactions in a state maximizing biomass production is, counter-intuitively, 50% smaller than in typical non-optimal states. More important, the in silico analysis identifies irreversible reactions as the root cause of this difference. This is so because the solutions of [1] in Box 1 under the constraints $v_j^\pm$ define a convex polyhedral solution region in the space of fluxes. A linear function in this region has an optimum at one of the vertices, where the constraints $v_j^- = 0$ of many irreversible reactions are binding, forcing those reactions and others coupled to them to be inactive. In the case of *E. coli*, it is estimated that over 70% of the metabolic reactions are irreversible under physiological conditions[19], and that some 20% others are effectively irreversible due to the presence of irreversible reactions in the same pathway[16]. Thus, as the metabolic system approaches an optimal state, some of these reactions will become inactive and trigger a cascade of inactivation that will propagate through other reactions in the network.

The observation that optimal states activate fewer reactions than non-optimal ones bears a strong relation to the rescuing effect of targeted reaction knockouts. The top candidates for rescue knockouts are precisely the reactions that are active in non-optimal states but predicted to be inactive in the desired state (see Fig. 2C). It is interesting to note that the larger the number of such reactions knocked out the stronger the strength of the rescue effect tends to be (see Fig. 3C, where this behavior is contrasted with the behavior observed for other forms of gene-gene interactions).

What is the theoretical maximum number of reactions that can be knocked out? This question admits a surprisingly simple and general answer for each fixed nutrient condition. The key observation that a reduced number of reactions are recruited by optimal states is not unique to *E. coli* or growth rate functions but is in fact a general behavior also found in other organisms and for many other linear functions of fluxes. Even more compelling, for optimal states under growth conditions, the number of active reactions was found to be ~300 for all typical linear functions and all four organisms—*H. pylori*, *S. aureus*, *E. coli*, and *S. cerevisiae*—studied in detail by Nishikawa et al.[16]. Then, the rule-of-thumb answer as to the maximum number of knockouts is that the strongest rescue would be achieved by removing all except ~300 reactions. The relevant question is to identify *which* 300 reactions to leave active, which does depend on the organism, the medium conditions, and the objective function to be maximized. Naturally, this is meant to be interpreted within the in silico model and with the caveat that the model ignores unknown side effects.

In all likelihood, these are instances of a general and ubiquitous behavior, which we are now only starting to understand. This behavior can be conceptualized in terms of a fitness landscape for a given objective function $\phi$, as illustrated schematically in Fig. 3A, B. A network (biological or not) that responds non-optimally to a perturbation will move to a state characterized by a value $\phi_1$ of this function that is smaller than its optimum $\phi_1^{opt}$. One can seek to force the network to a state closer to this optimal by eliminating states corresponding to small $\phi$. By doing so, the new best state available may change to a value $\phi_2^{opt}$ smaller than the original optimum $\phi_1^{opt}$. Nevertheless, the fitness of the system will still be improved if this shifts the system to a new state with fitness $\phi_2$ larger than $\phi_1$. Such an improvement is expected to be possible in general by constraining (rather than augmenting) the dynamics on the network and, in particular, by removing nodes and edges as in the case of reaction knockouts. This intervention only exploits resources and states originally available in the system. Accordingly, I refer to this intervention as a *restraint-based approach* to control the network response.

In the specific case of cellular metabolism, if the reaction rates are constrained to (but not less than) their optimal values[5], then the optimal does not change and $\phi_2^{opt}$ is guaranteed to remain equal to $\phi_1^{opt}$ (Fig. 3B, blue curve). By bringing the individual fluxes closer to the optimal fluxes we expect to bring the equilibrium of the entire system, and by implication the metabolic objective, closer to its optimum. It should be emphasized that because this is a control to a response, not all states corresponding to small $\phi$ have to be eliminated for this to be achieved. In fact, states in which most fluxes are zero or close to zero always exist simply because they are always solutions of Eq. [1] under the constraints $v_j^\pm$ (Box 1).

The reason I emphasize reaction knockouts is not a mere caprice. The reciprocal would be reaction up-regulation which, although more logical as a way to direct resources to the desired pathways, are not easy to be implemented in practice because the regulation of one reaction often requires the coordinated fine-tuned regulation of various other reactions[20]—but see Ref. (21) for a discussion on expression systems. The insertion of a new gene to cover for a fault can be an exception to this since the rest of the cellular machinery, if intact, can coordinate such response as in the original system. (Fitness advantage may also be conferred by rewiring of the network circuitry at the signaling or regulatory level[22], although issues related to the stability and robustness of such changes have just started to be elucidated[23,24].) Gene expression, in particular, is not guaranteed to correlate with reaction activity. This illuminates another property of the metabolic network that makes this approach useful in practice. If the metabolic network had the property of activating most or all reactions in optimal states, one could still conceive a restraint-based approach to enhance the response of the network by *partially* knocking out (i.e., knocking down) the genes and hence reactions that are found to be over-expressed in the observed non-optimal states (cf. Fig. 2C). However, this would require determining the exact relation between gene expression and reaction flux, whereas full knockouts are properly defined without the need of such information.

## Design of Antagonistic Drug Combinations

Two drugs may exhibit no interaction, interact synergistically, or interact antagonistically. The latter includes cases in which the strength of the two-drug combination is weaker than one of them alone. In the case of antibiotics, for example, in the presence of one antibiotic the addition of a second antibiotic would increase rather than decrease bacterial growth. Kishony's lab has recently demonstrated that, counter-intuitively, such suppressive drug combinations can select against resistant strains[25]. This was demonstrated using an ingenious experiment with sub-lethal doses of the antagonistically interacting antibiotics doxycycline and ciprofloxacin. By measuring the effect of this combination in direct competition between doxycycline-sensitive and doxycycline-resistant *E. coli*, they could identify a region in the concentration diagram where the growth rate of the sensitive strain overcomes that of the resistant mutants. This is so because mutations conferring resistance to doxycycline are effectively equivalent to a reduction in the concentration of this drug, which increases the effect of ciprofloxacin (see Fig. 4A).

Similar results have been previously obtained by Blagosklonny using the anticancer drugs doxorubicin and paclitaxel in HL60 human leukemia cells[26]. The combination was shown to select against doxorubicin-resistant cells due to the suppressive interaction between doxorubicin and paclitaxel. Resistance is in this case associated with the expression of transporter MRP1 and consequent efflux pump of doxorubicin. This efflux and resulting reduction in doxorubicin concentration elucidates how resistance equates to a reduction of this drug in the concentration diagram of Fig. 4A (horizontal arrow). Furthermore, recent evidence indicates that, when compared to synergistic interactions, antagonistic drug combinations are less likely to lead to the evolution of resistance in the first place[27].

It is therefore of much interest to develop a rational approach to identify drug combinations that exhibit antagonistic and, in particular, suppressive interactions that can bias selection against resistance. As suggested above, this would be important not only in the development of antibiotics but also in the development of anticancer drugs. Many different mechanisms may underlie suppressive drug interactions, but given that drug interactions are analogous to genetic epistatic interactions, synthetic rescues certainly appear as a very promising candidate to be explored.

Figure 4B schematically shows the analogue of Fig. 4A for synthetic rescue gene pairs, establishing a very important relation that has not been previously appreciated. In this case the axes represent the level to which the gene expression is suppressed, ranging from no intervention to full knockout. In this example, the knockout of gene A will select against strains carrying gene B, since the knockout of the latter can rescue the knockout of the former. This is derived under the simplifying assumption that the resulting growth rate does not depend on the order of the gene removals, or that they are implemented simultaneously. In this argument, gene B could be replaced by a function gained by resistance that is difficult to be targeted directly, such as transporter MRP1 in the case of cancer, or any other cellular function that distinguishes the target cells from "normal" cells. To select against these cells, the question is then reduced to the task of identifying gene knockouts that would (hypothetically) be rescued by the inactivation of that function. Drugs that would exploit these targets would thus interact antagonistically through the effect they cause on the physiology of the cell as opposed to direct chemical interactions between the drugs. In the whole-cell scale, the same can be conceived when gene A is replaced by a set of multiple genes or proteins and the gene knockout is replaced by a continuous modulation of these components.

The relation between synthetic rescues and antagonistic drug interactions is further strengthened by the very recent demonstration of the molecular mechanism underlying the suppression of DNA synthesis-inhibiting antibiotics by protein synthesis-inhibiting antibiotics[28]. In this case, a DNA synthesis inhibitor such as trimethoprim results in a rate of protein synthesis above the optimal value for maximum growth, which can be partially remediated by adding a translation inhibitor such as spiramycin. The above-optimal protein synthesis—and the resulting suppressive interaction between the antibiotics trimethoprim and spiramycin—is caused by the non-optimal response of ribosomal gene expression to the presence of DNA stress imposed by the DNA synthesis inhibitor[28]. This mechanism affords direct parallels with the non-optimal response to genetic perturbations that underlies synthetic rescues[16].

Specifically in the context of cancer, the possibilities this may open for combinatorial therapeutics are very much in line with the recent shift in anticancer drug development from cytotoxic drugs to more specific agents that modulate proteins associated with cancerous states[29]. On the other hand, a recent study of approved drugs showed that drugs acting on single targets appear to be the exception rather than the rule[30]. But lack of selectivity is not necessarily an undesirable property[31] given that complex diseases, such as cancer, may not be effectively treated by modulating single targets, a hypothesis that is further corroborated by the fact that many of the most effective drugs have this property. In the case of antibiotics, synthetic lethality provides a basis for why this would be the case, and analogous effects may be at work in the case of anticancer drugs believed to act on multiple signaling proteins, such as imatinib and sunitinib[32]. It is therefore expected that the combination of two or more antagonistically interacting single- or multi-target drugs will lead to a rational new approach to explore these possibilities, with its own advantages and challenges. Of utmost interest for this exploration is the need to determine the impact of antagonistic drug combinations on normal cells of the host organism vis-à-vis the impact of drug combinations exhibiting synergistic interactions or no interactions.

## Implications and Applications

In addition to their role in antagonistic drug interactions, synthetic rescues and related concepts have the potential to provide new insights into numerous outstanding problems, including:

**1. Lethality vs. Essentiality.** It has been largely assumed in the literature that a gene whose knockout is lethal is necessarily essential for growth, but the occurrence of synthetic rescues shows that this is generally true *only* if no other genes are concurrently knocked out or knocked down. Thus, the notion of gene essentiality has to be distinguished from the notion of lethality even when the environmental conditions are kept unchanged. This is illustrated in Fig. 5, where I reproduce some predictions for *E. coli* in glucose minimal medium. In this medium, gene *pgk* is essential because the knockout mutants are not able to produce biomass regardless of the expression level of the other genes. The knockout of gene *pfkAB*, on the other hand, is lethal but this gene is not essential because the knockout mutants are predicted to recover the ability to produce biomass upon the concurrent knockout of genes *lpdA*, *gpt*, *gadB*, *gadA*, *tynA*, *aceA*, *gltP*, *gltS*, and *pat*. (This example also illustrates that synthetic rescues may require multiple gene knockouts[5].) Therefore, while the knockout of an essential gene is guaranteed to be lethal, the converse—that lethal knockouts would correspond to essential genes—is generally not true. Moreover, while both essentiality and lethality depend on the environmental conditions, only the latter depends on the state of the system prior to the gene knockout. Note that this distinction between lethality and essentiality does not follow from usual compensatory mutations, since those tend to add function to the system, nor from alternative definitions of 'synthetic rescues' that include growth recovery caused by mutations other than knockouts[33].

**2. Dependence on Initial Conditions.** The impact of an environmental or genetic perturbation depends not only on the environmental and genetic background but also on the level of adaptation of the cells, as already suggested by the hypothesis of minimization of metabolic adjustment[34]. More striking, whether a perturbation will be *lethal or not* depends critically on the specific pre-perturbation metabolic state of the cell. This otherwise subtle dependence may explain part of the current disagreements between different knockout experiments conducted under apparently similar conditions. For example, for *E. coli* K-12 in rich media, 240 out of 303 "essential gene" candidates identified in the Keio collection[35] had previously been tested by Gerdes et al.[36] and PEC collection[37], but only 60% of them were found to be essential in both of these previous studies [Natali Gulbahce, private communication]. If one considers genetic and environmental changes as inputs and growth or other integrated function as an output, the dependence on the internal state of the cell determined by the previous history of the strain is in many aspects analogous to hysteresis in physical systems. This is important given that numerous publications on growth experiments (including some of the most inspiring ones) do not uniquely define the initial state of the cells. Overnight growth in rich medium, for example, is analogous to have a magnetized material relaxing in a magnetic field for a certain period of time, which by itself does not say what state the system reached at the end of the process. This is partially related to the fact that, in the case of two or more gene knockouts, the order and timing of the knockouts may matter.

**3. Gene Dispensability Conundrum.** The transient activation of latent metabolic reactions following a perturbation tends to involve a large number of reactions that are asymptotically inactive both before and after the perturbation. That is, the transient activity includes much more than the union of the optimal sets of reactions that would be recruited before and after the perturbation (up to 2 times larger for *E. coli* in minimal glucose medium[16]). Thus, the number of reactions and hence genes that are active under variable conditions can be significantly larger than under standard laboratory conditions. This provides natural new hypotheses to address the longstanding problem of gene dispensability—the observation that the knockout of numerous genes have negligible impact on growth under standard laboratory conditions[38,39] and, to a lessen degree, even when alternative nutrient conditions are considered[40] (but see also Ref. (41), which addresses the impact of chemical stress). One such hypothesis is that the presence of latent (otherwise dispensable) pathways facilitate adaptation and hence confer competitive advantage under variable conditions, which is a possibility that has not been fully explored in previous studies. In particular, it may be the case that apparently neutral mutations are not neutral after all[42]. This would certainly help explain why bacteria and other highly optimized organisms activate latent pathways in the first place. However, there is an obvious problem with this argument, namely that the deletion of latent reactions has been found to rescue rather than aggravate growth defect[5,16]. This would make them not only dispensable but also undesirable. This alone does not rule out the possibility that latent pathways may offer some benefit not captured by in silico models, but it does make the problem more interesting and increases the plausibility that each of these reactions is permanently needed for growth in at least one evolutionarily relevant condition not yet identified. But why are latent pathways activated when they are not permanently needed after the perturbation? One possibility is that they create plasticity by generating a library of possible states from which the regulatory system can choose mainly by down- rather than up-regulating reactions. This possibility is consistent with the hypothesis that suboptimal states serve as standby states in variable environments, as the suboptimal growth of wild-type organisms even in their preferred carbon source seems to suggest[43,44].

**4. Tolerance to Environmental Stress.** Synthetic rescue interactions improve fitness at the expense of robustness. The rescued organisms will grow faster in the environmental conditions under consideration but will (presumably) lose flexibility to adapt to other conditions. For many applications, however, such as in the microbial production of compounds of industrial interest, this is a secondary problem. The primary problem is to improve tolerance to the specific industrial environment (e.g., tolerance to specific ethanol and oxygen concentration in the case of yeast). Therefore, the systematic study of synthetic rescues may add a new dimension to microbial optimization efforts, a field in which gene knockouts have been previously used to optimize production of specific metabolic compounds as byproducts of adaptation-induced growth optimization[45,46].

## Other Systems and Processes

Studies of both human[47] and animal[48] cerebral cortices have shown that a significant fraction of the synapses created in the first stage of brain development is eliminated before adulthood. In the frontal human cortex, the maximum synaptic density takes place at 1-2 years of age and is up to 50% higher than the adult mean[49]. Like in the case of latent pathways, this transient is believed to generate plasticity, in this case by creating an anatomical substrate for future brain development. As such, synapse elimination is part of normal brain development. Interestingly, similar effect plays a role in the theoretical design of neural networks for pattern-recognition tasks. In the latter context, it has been shown that the selective elimination of network connections can improve the performance of already good networks while significantly reducing the number of parameters[50]. This is in fact the basis of an elegant technique for neural network optimization known as optimal brain damage[50].

Back to the real brain, another puzzling observation concerns the paradoxical effect of lesions[51]. It has been found, in particular, that while certain unilateral brain lesions can lead to reduced spatial attention, the addition of a second lesion to the other hemisphere often leads to partial restoration of the lost function[52]. Is it a mere coincidence that such disparate systems exhibit behavior so strikingly similar to the one observed in metabolic networks?

Possibly, but it is more plausible to admit that these systems have common properties owing to their underlying network structure. With that in mind, one can expect that a restraint-based approach of the type discussed above can be used to rescue and control numerous networked systems, including non-biological ones. In traffic control, for example, congestion pricing is an efficient method to reduce congestion by surcharging users in periods of peak demand[53]. This is effectively equivalent to constraining high-demand roads by stimulating users to re-route or reschedule their journey in order to shorten everyone's travel time. Control of overload cascading failures in power grids is another illuminating example. In power grids, the flows can be re-routed to minimize overloads by exploiting the local conservation of currents at each power station. By comparison, in metabolic networks, a similar conservation law—the conservation of mass—is at work in each reaction and underlies the re-routing of fluxes. In this analogy, knockouts and changes of medium conditions are tantamount to load shedding and dispatch of power generation, respectively. In fact, the original findings on synthetic rescues in metabolic networks were partially inspired by a method introduced in Ref. (54) to control cascading failures in complex networks. It is therefore not entirely surprising that these disparate systems exhibit similar phenomena.

It is also tempting to compare synthetic rescues with Olson's theory[55] that "loss of gene function may represent a common evolutionary response of populations undergoing a shift in environment and, consequently, a change in the pattern of selective pressures." Evidence for this theory is provided, for example, in the recently observed high frequency of nonsense single-nucleotide polymorphisms (SNPs) in human populations, which suggests that truncation and even inactivation of some specific proteins have been advantageous in recent human evolution[56]. In a population of 1,151 individuals, 99 genes where found with both copies inactivated by nonsense SNPs in at least one individual. Such loss of gene function can occur in isolation. However, the existence of synthetic rescues suggests that the inactivation of certain genes could be compensatory, to the extent that they would be selected for after the inactivation of another gene even when they would be selected against in isolation. Therefore, in addition to the possibility identified by Olson's theory, in which loss of function is driven by environmental changes, synthetic rescues indicate that loss of gene function may also be driven by a previous deleterious genetic modification. Empirical data does not exclude this possibility. In fact typical healthy individuals are found to have variations due to nonsense SNPs in tens of their genes[56].

## Conclusions

As argued here, the response of a decentralized complex network can be largely controlled and optimized by constraining its structure and/or dynamics or, more generally, the resources available in the system. Moreover, biological networks appear to have evolved to operate with a mechanism of selection a posteriori. This is plausibly the case for cellular metabolism, which transiently activates latent pathways whenever adaptation to a new condition is called for, generating a library of states it can select from via down-regulation. This transient activation also illustrates how unlikely it is that a network will optimize any given objective function in the absence of adaptation. What is more, there are cases, such as those associated with lethal perturbations, in which adaptation alone cannot lead to the optimization of the (natural or human selected) function of interest. But the study of synthetic rescues[5] provides a clear procedure to identify compensatory perturbations based on constraining the reactions that are over-activated (or run in the opposite direction) when compared to the desired optimal state.

These compensatory perturbations are by no means evident from the structure of the network, highlighting the importance of a modeling and experimental framework that can account for both the nonlinear and the system-level nature of the network response to perturbations. Moreover, because synthetic rescues involve the inactivation of two (and often more) genes, it is imperative to develop methods to systematically study the response of cellular networks to multiple perturbations. The expected results of this research are very promising, as they will address outstanding questions about collective gene interactions and potentially lead to new approaches for drug development that, like in the story of Lorenzo's oil, involve interactions that challenge common sense. The latter can have direct implications for medical research related to metabolic diseases and, combined with current studies on antagonistic drug interactions[8], lead to a new paradigm to address drug resistance in antibiotic and cancer research. This is only possible, however, if we can understand how the cellular network responds to multiple perturbations.

Therefore, whether you call it systems biology, integrative biology, or network biology, it is clear that we need more of it.

**Acknowledgments:** The author is supported by the National Science Foundation under Grant DMS-0709212, a Sloan Research Fellowship, and the National Cancer Institute under Grant 1U54CA143869-01, and graciously thanks Luciana Zanella for illuminating discussions.

**BOX 1:** Main elements of the computational modeling of cellular metabolism.

Metabolic networks can be modeled using in silico reconstructions that account for biochemical and transport reactions, biochemical species, and reaction-enzyme-gene relationships[57]. The state of the network is described by the vector $\mathbf{v} = (v_j)$ of the reaction fluxes. Complications due to unknown regulatory mechanisms and kinetic parameters are avoided by focusing on steady-state dynamics and describing the response to perturbations as transitions between steady states. The steady-state approximation is generally appropriate to describe the short-term behavior of individual cells as well as the average long-term behavior of large populations of cells. For a network with $n$ fluxes and $m$ species, the steady-state solutions are determined by

$$\sum_{j=1}^{n} S_{ij} v_j = 0, \qquad i = 1,\ldots,m, \qquad [1]$$

where $\mathbf{S} = (S_{ij})$ is the stoichiometric matrix accounting for the network structure. This matrix typically consists of hundreds or thousands of reactions and a large but smaller number of biochemical species, rendering Eq. [1] underdetermined. The individual fluxes are further bounded as $v_j^- \leq v_j \leq v_j^+$ to model, for example, constraints imposed by thermodynamics, availability of nutrients, and maximum reaction rates[19,58]. In particular, $v_j^\pm = 0$ is used for uptake reactions of nutrients not available in the medium and $v_j^- = 0$ for reactions that are irreversible. The resulting system is still underdetermined, as expected since cell regulation is not explicitly incorporated into the model. Phenomenological methods such as flux balance analysis (FBA)[59,60], minimization of metabolic adjustment (MOMA)[34], and regulatory on/off minimization (ROOM)[15], can then be used to implement biological hypotheses that predict metabolic behavior by selecting one out of the multiple solutions of Eq. [1]. FBA identifies a solution that optimizes a linear function of fluxes, such as biomass production (and hence growth rate) in the case of single-cell organisms well adapted to their environment. MOMA and ROOM model responses to perturbations, such as a gene knockout implemented by setting $v_j^\pm = 0$ for the corresponding reactions. MOMA provides a solution compatible with the constraints imposed by the perturbation that is closest to the original metabolic state in the space of fluxes, while ROOM minimizes the number of significant flux changes. Thus, a perturbed metabolic network will generally depart from its optimal states.

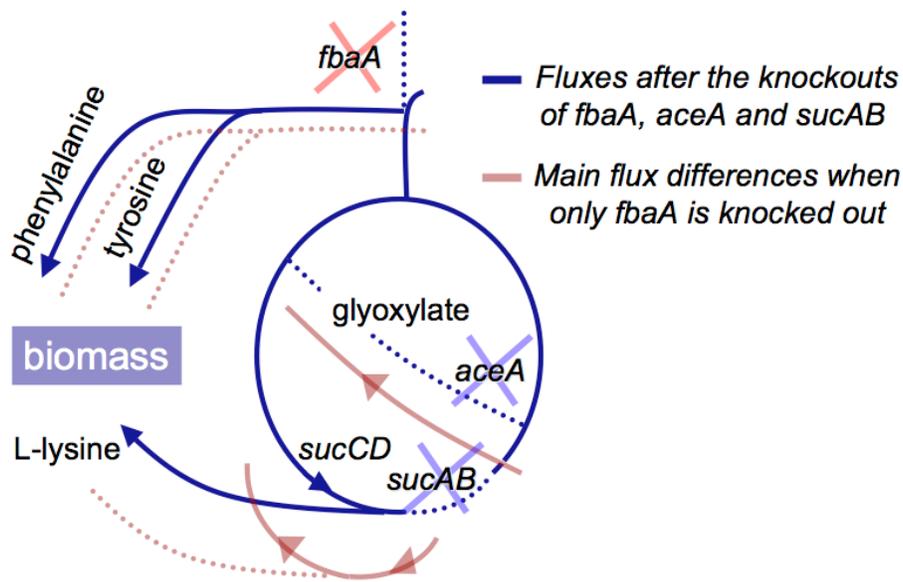

**Figure 1.** Example of synthetic rescue in the TCA cycle of *E. coli* K12 MG1655 fed arabinose. Following the deletion of gene *fbaA* the cell is unable to produce phenylalanine, tyrosine and L-lysine, which are components of the biomass. However, the deletion of gene *aceA*, gene *sucAB*, or both, as shown here, restores the ability of the *fbaA*-deficient cell to produce biomass because the inactivation of the *aceA*- and *sucAB*-catalyzed reactions increases the fluxes of other reactions involved in the production of biomass components. All relevant fluxes are shown after the rescue knockouts (blue lines) and the fluxes corresponding to the main flux changes are also shown before the rescue knockouts (red lines), where nonzero fluxes are represented by continuous lines and zero fluxes by dotted lines. In a state that maximizes biomass production, the reactions removed by the rescue knockouts would not be active and other reactions—such as the one catalyzed by gene *sucCD*—would run in the opposite direction, which illustrates that the rescue perturbations effectively correct for the suboptimal response to the primary perturbation. Indeed, while previous studies have considered the removal of competing pathways to increase microbial production of certain chemicals, synthetic rescues demonstrate that gene knockouts can be used to control the *response* to perturbations. The inability to produce biomass following the knockout of gene *fbaA* and the recovery of this ability upon the rescue knockouts is entirely due to the response of the network, in this case modeled using MOMA, since the FBA solutions—those that maximizes growth—are not altered by the rescue knockouts (see Box 1). (Figure adapted from Motter et al.[5], based on simulations of the entire metabolic network.)

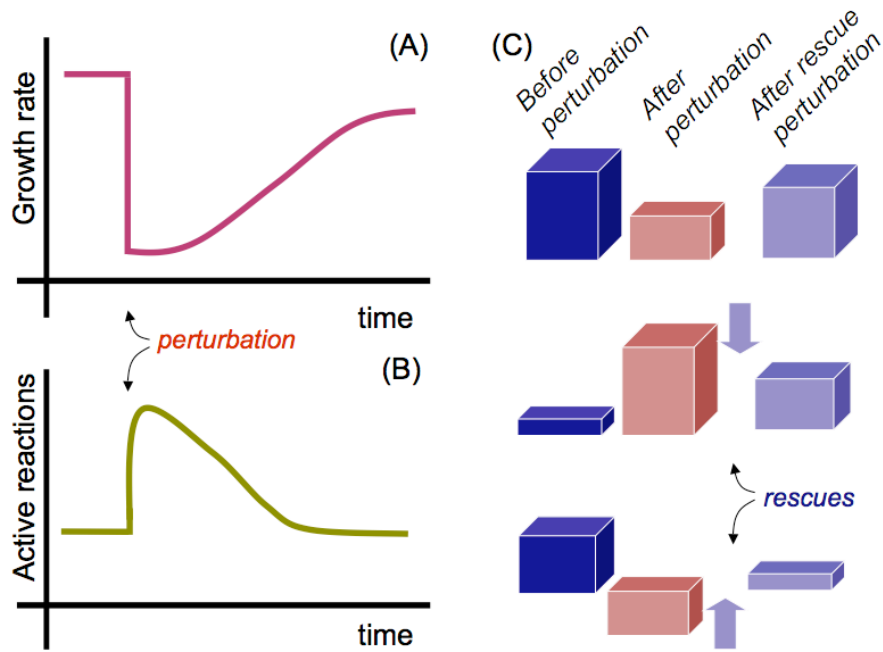

**Figure 2.** Impact of perturbations on cellular growth and metabolic reaction activity. (A,B) A transient decrease in growth (A) is often accompanied by a transient activation of latent reactions (B). After adaptive evolution, the growth and reaction activity can become strikingly similar to the original one. (C) Reaction-flux changes over time scales shorter than those relevant for adaptive evolution, which also apply to cases in which adaptive evolution is not possible (e.g., in the presence of a lethal perturbation). The reactions can be divided into three groups: those whose fluxes will decrease (top red), those whose fluxes will increase (middle red), and those whose fluxes will change direction (bottom red) in response to the growth-suppressing perturbation. The fluxes in all three groups can be partially or completely corrected by rescue knockouts implemented in the second and/or third group of reactions (middle and/or bottom light blue).

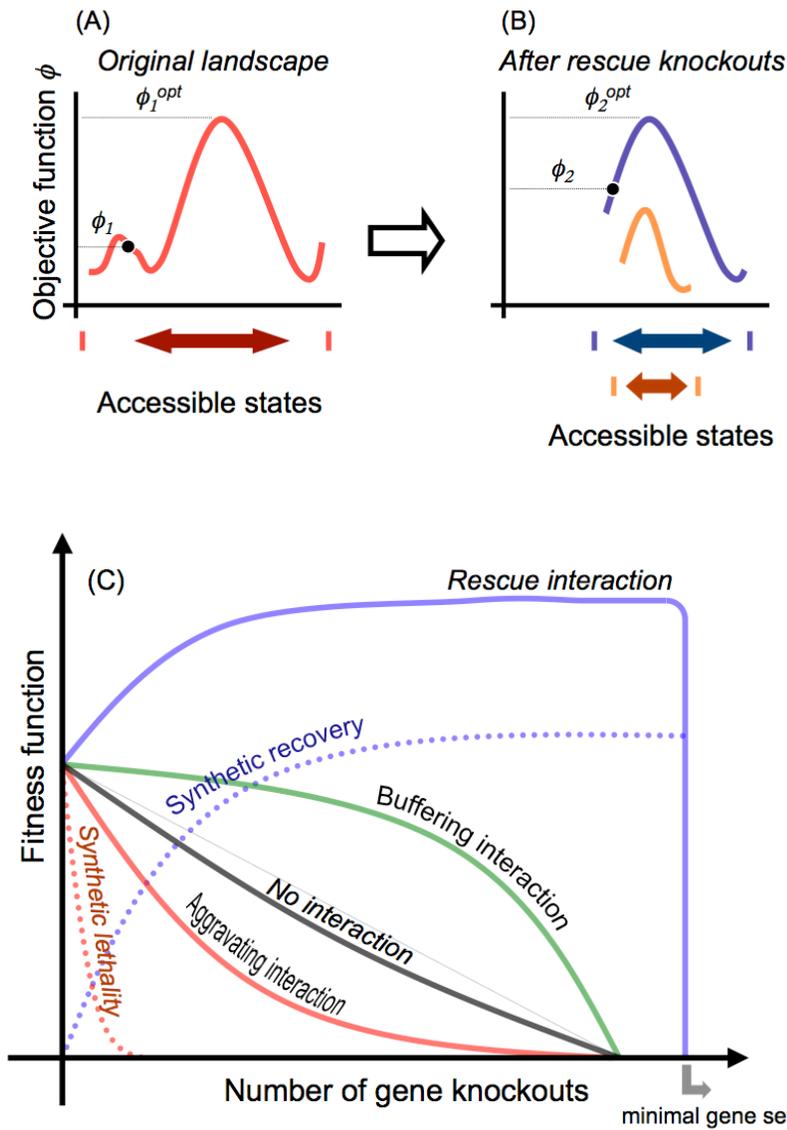

**Figure 3.** Impact of restraint-based perturbations and their relation to gene-gene interactions. (A,B) Example fitness landscape in which the system is found to be in a non-optimal state (A) and moves to a state that is closer to optimal after the elimination of some of the accessible states (B). While restraint-based perturbations generally represent a compromise between eliminating undesirable responses and keeping the objective function of the available states as high as possible, as illustrated by the yellow curve, the synthetic rescues discussed in the text are guaranteed not to affect the optimal, growth-maximizing states, as illustrated by the blue curve. (C) Rescue interactions relative to other epistatic gene-gene interactions for a common gene-deficient mutant (continuous curves), where the no-interaction case would correspond to a straight line in a log-linear representation. In the same way *synthetic lethality* can be regarded as an extreme case of aggravating interaction, *synthetic recovery* (dotted blue curve)—where a nonviable gene-deficient strain is rescued by one or more gene knockouts—can be regarded as an extreme case of rescue interaction (continuous blue curve). Note that classic epistatic interactions (aggravating, non-interacting, and buffering) are determined as averages over random gene knockouts, whereas rescue interactions are conceived as *targeted* gene knockouts, indicating that they can in general coexist. The rescue interaction curve is characterized by a fast increase for a small number of knockouts, followed by a plateau where additional knockouts have no effect because those reactions have been inactivated by an inactivation cascade triggered by the first ones, followed by a sudden drop in fitness. The latter has been predicted to correspond to a remaining set of approximately 300 unique metabolic reactions for the several organisms and objective functions considered by Nishikawa et al.[16] and is in fact a close approximation to the minimal genome set.

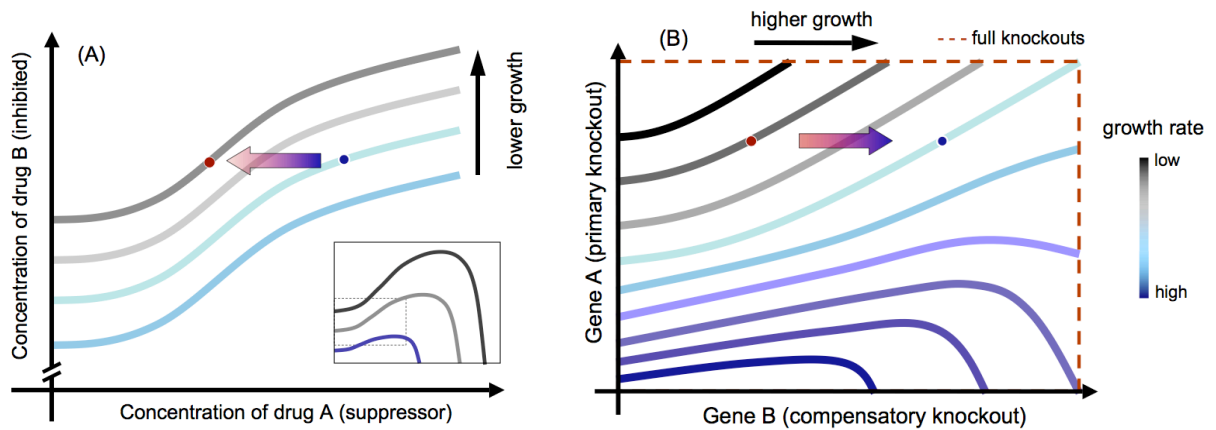

**Figure 4.** Curves of constant growth rates for antagonistic drug interactions and synthetic rescues. (A) Illustration of antagonist interaction in which drug A (e.g., doxycycline or doxorubicin) suppresses the activity of drug B (e.g., ciprofloxacin or paclitaxel, respectively). Color lines represent equal effect of the two-drug combination on growth. The horizontal arrow indicates a change in growth rate due to resistance to drug A, illustrating that, under the concurrent action of drug B, a drug A-resistant strain will have *lower* growth rate than a non-resistant strain. The inset shows the full diagram, where the main figure corresponds to a portion of the dotted box region (adapted from Chait et al.[25]). (B) Illustration of a synthetic rescue interaction in which the knockout of gene A is rescued by the knockout of gene B, where the axes range from no intervention (bottom, left) to full knockouts (top, right). The horizontal arrow indicates a change in growth caused by the knockout of gene B, illustrating that the double knockout strain will have *higher* growth rate than the gene A-deficient strain.

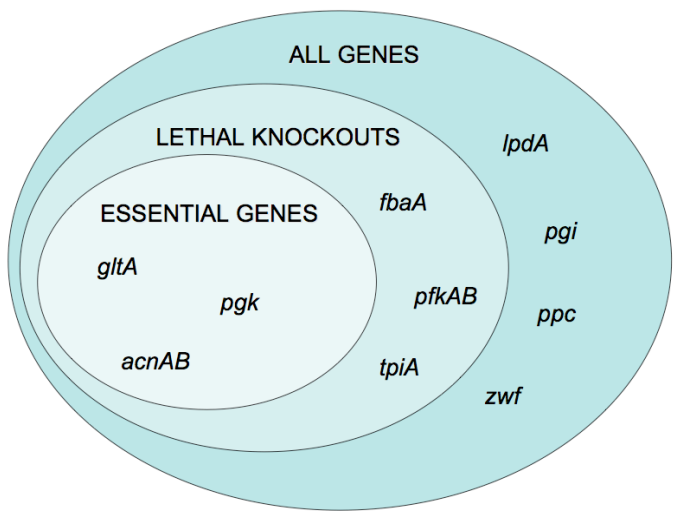

**Figure 5.** Lethality-essentiality chart for *E. coli* K12 MG1655 fed glucose. Genes can be naturally organized into three groups: those that are essential, those that are not essential but whose knockouts are lethal, and those whose knockouts are not lethal. Mutants generated by the knockout of a gene in the intermediate group, such as *tpiA*, can in general be synthetically rescued by the knockout of one or more additional genes. The examples given on the chart are consistent with in silico and experimental observations (Ref. (5) and references therein).